\documentclass[aps,prd,onecolumn,superscriptaddress]{revtex4}

\usepackage{graphicx}
\usepackage{times}
\usepackage{dcolumn}
\usepackage{color}

\newcommand{\be}{\begin{equation}}
\newcommand{\ee}{\end{equation}}
\newcommand{\bea}{\begin{eqnarray}}
\newcommand{\eea}{\end{eqnarray}}

\newcommand{\bk}{{\bf k}}

\newcommand{\bs}{{\bf s}}
\newcommand{\cs}{{\cal S}}

\newcommand{\deltar}{\delta_{\rm recon}}

\begin{document}

\title{Shot noise and reconstruction of the acoustic peak}

\author{Martin White}
\affiliation{Departments of Physics and Astronomy, 601 Campbell Hall,
University of California Berkeley, CA 94720}

\date{\today}

\begin{abstract}
We study the effect of noise in the density field, such as would arise from
a finite number density of tracers, on reconstruction of the acoustic peak
within the context of Lagrangian perturbation theory.
Reconstruction performs better when the density field is determined from
denser tracers, but the gains saturate at
$\bar{n}\sim 10^{-4}\,(h\,{\rm Mpc}^{-1})^3$.
For low density tracers it is best to use a large smoothing scale to define
the shifts, but the optimum is very broad.

\end{abstract}


\maketitle
\twocolumngrid

\section{Introduction}

Baryon acoustic oscillations (BAO) in the baryon-photon fluid provide a standard
ruler to constrain the expansion of the Universe and have become an integral
part of current and next-generation dark energy experiments \cite{EisReview05}.
These sound waves imprint an almost harmonic series of peaks in the power
spectrum $P(k)$, corresponding to a feature in the correlation function
$\xi(r)$ at $\sim$100 Mpc, with width $\sim 10$\% due to Silk damping
\cite{PeeYu70,SunZel70,DorZelSun78,Eis98,MeiWhiPea99,ESW}.
Non-linear evolution leads to a damping of the oscillations on small scales
\cite{Bha96,MeiWhiPea99} (and a small shift in their positions
\cite{ESW07,CroSco08,Mat08a,Seo08,PadWhi09}),
\begin{equation}
  P_{\rm obs}(k) = b^2 e^{-k^2\Sigma^2/2} P_L(k) + \cdots
  \cdots
\label{eq:processed}
\end{equation}
where we have assumed a scale-independent bias, $b$, and left all broad
band and mode-coupling features implicit in the $\cdots$.
The damping of the linear power spectrum (or equivalently the smoothing of
the correlation function) reduces the contrast of the feature and
the precision with which the size of ruler may be measured and is given by
the mean-squared Zel'dovich displacement of particles,
\begin{equation}
  \Sigma^2 = \frac{1}{3\pi^2} \int dp\ P_L(p) \qquad .
\label{eq:sigmal}
\end{equation}

In \cite{ESSS07} a method was introduced for reducing the damping,
sharpening the feature in configuration space or restoring the higher
$k$ oscillations in Fourier space.
This procedure was studied in \cite{PadWhiCoh09,NohWhiPad09} using
Lagrangian perturbation theory.
In this brief note we generalize these treatements to show how the effects of
noise in the density field, arising for example from the finite number density
of tracers, affects reconstruction.  We shall concentrate on the broadening of
the peak, and refer the reader to \cite{PadWhiCoh09,NohWhiPad09} for details,
discussion and notation.

\section{Reconstruction with noise}

The prescription of \cite{ESSS07} begins by smoothing the observed density field
to filter out high $k$ modes: $\delta(\bk)\rightarrow \cs(k)\delta(\bk)$.
We shall take $\cs$ to be Gaussian of width $R$.  From the smoothed field
the negative Zel'dovich displacement is computed
$\bs(\bk)\equiv -i(\bk/k^2)\cs(k)\delta(\bk)$.
Then the objects are shifted $\bs$ to form the ``displaced'' density field,
$\delta_d$, and an initially spatially uniform grid of particles is also
shifted to form the ``shifted'' density field, $\delta_s(\bk)$.
The reconstructed density field is defined as $\deltar\equiv\delta_d-\delta_s$,
and to lowest order it is equal to the linear density field
\cite{ESSS07,PadWhiCoh09,NohWhiPad09}.
The non-linear damping is however modified from $\exp[-k^2\Sigma^2/2]$ to
\cite{PadWhiCoh09,NohWhiPad09}
\begin{eqnarray}
  D(k)&\equiv&
  \cs^{2}(k) e^{-\frac{1}{2} k^{2} \Sigma_{ss}^{2}}
  +  [1-\cs(k)]^{2} e^{-\frac{1}{2} k^{2} \Sigma_{dd}^{2}} \nonumber \\ 
  &+& 2 \cs(k) [1-\cs(k)] e^{-\frac{1}{2} k^{2} \Sigma_{sd}^{2}}  \,\,.
\label{eq:damptransform}
\end{eqnarray}
with $\Sigma_{ss}$ and $\Sigma_{dd}$ defined as integrals over the linear
power spectrum, $P_L$, (see below) and
$\Sigma_{sd}^{2}\equiv (1/2)\left(\Sigma_{ss}^{2}+\Sigma_{dd}^{2}\right)$.

If we assume there is a contribution, $\delta_N$, from noise we find
$\deltar$ is unchanged to lowest order.  However the damping scale is
modified.  Following \cite{PadWhiCoh09} we find
\begin{equation}
  \Sigma_{ss}^{2} \to  \frac{1}{3\pi^2} \int dp\ \cs^2(p)
  \left[ P_L(p) + P_N(p) \right]
\end{equation}
where $P_N$ is the power spectrum of $\delta_N$ and
\begin{equation}
  \Sigma_{dd}^{2} \to  \frac{1}{3\pi^2} \int dp\ \left[1-\cs(p)\right]^2
  P_L(p) + \cs^2(p) P_N(p) \, ,
\end{equation}
which reduce to the expressions of \cite{PadWhiCoh09,NohWhiPad09} as $P_N\to0$.
For Poisson shot-noise we expect $P_N=b^{-2}\bar{n}^{-1}$ for tracers with
number density $\bar{n}$ assuming linear bias $b$.
These equations present the generalization of the treament in
\cite{PadWhiCoh09,NohWhiPad09} to include shot-noise.

\section{Results}

\begin{figure*}
\begin{center}
\resizebox{2.8in}{!}{\includegraphics{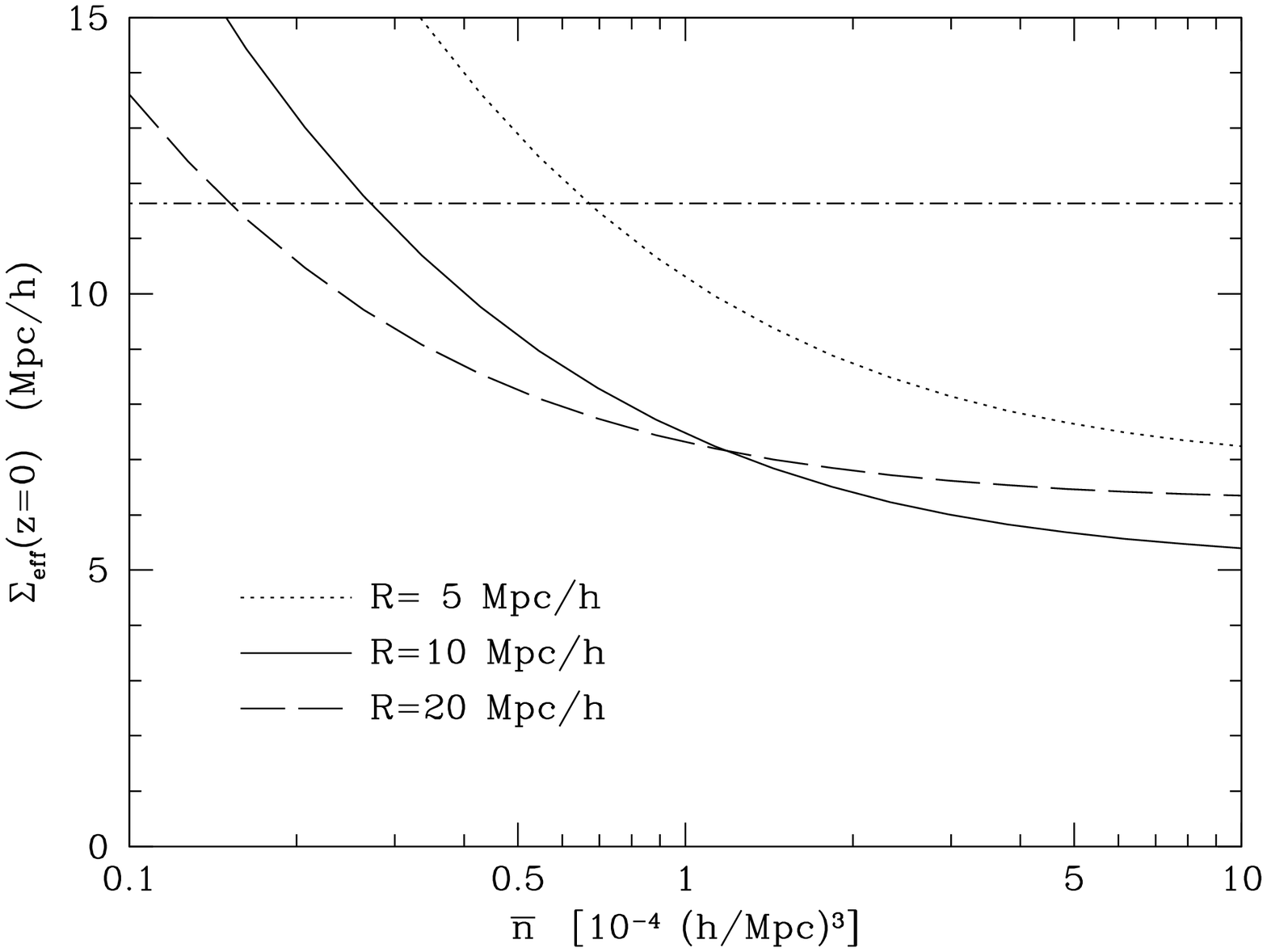}}
\resizebox{2.8in}{!}{\includegraphics{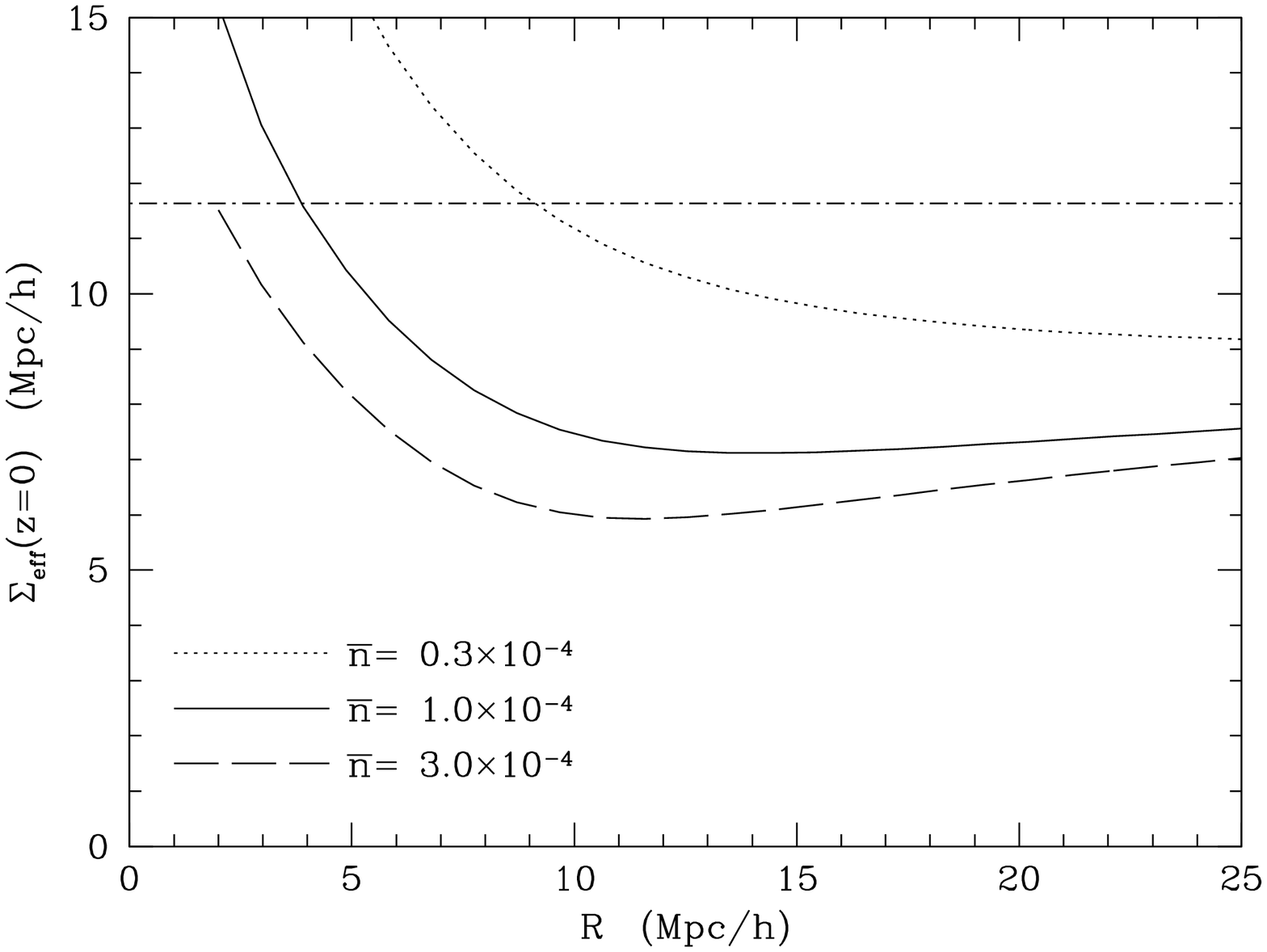}}
\end{center}
\vspace{-0.2in}
\caption{(Top) The damping scale at $z=0$, $\Sigma_{\rm eff}$, after
reconstruction as a function of the number density of tracers, $\bar{n}$,
assuming $b=1$.
The horizontal dot-dashed line indicates the Silk scale, or the intrinsic
width of the acoustic peak, for our cosmology.
(Bottom) As above but as a function of smoothing scale, $R$.}
\label{fig:SigvsNR}
\end{figure*}

One method to forecast the effect of this noise on cosmological parameters
constrained by BAO is to replace the Gaussian damping of
Eq.~(\ref{eq:processed}) with
Eq.~(\ref{eq:damptransform}) in the computation of the Fisher matrix for
the acoustic scale $s$.  For example, in spherical geometry \cite{SeoEis07}
\begin{equation}
  \sigma^{-2}_{\ln s} = \frac{V_{\rm survey}}{2}\int\frac{d\bk}{(2\pi)^3}
  \left[\frac{\partial P/\partial\ln s}{P+\bar{n}^{-1}}\right]^2
\label{eq:Fisher}
\end{equation}
with
\begin{equation}
  P \propto D(k)\frac{\sin ks}{ks} e^{-k^2\Sigma_{\rm Silk}^2/2} + \cdots
\end{equation}
where $\Sigma_{\rm Silk}$ is the Silk damping scale and $\cdots$ refers to
terms independent of $s$ \cite{SeoEis07}.  The effects of shot-noise show up
in the increased damping of the higher harmonics of the signal and the
increase in the variance per $\bk$ mode
(the denominator in Eq.~\ref{eq:Fisher}).

However, almost as much intuition can be gained by approximating $D(k)$ as
a Gaussian and asking how the effective damping depends on $P_N$.  To this
end we define an ``effective'' $\Sigma$ from the value of the damping at
$k_{\rm fid}=0.2\,h\,{\rm Mpc}^{-1}$.

Figure \ref{fig:SigvsNR}, top, shows how $\Sigma_{\rm eff}(z=0)$
depends on $\bar{n}$ for a $\Lambda$CDM model with $\Omega_m=0.25$.
Note that reconstruction improves for higher number density tracers, but
the gains saturate above approximately $10^{-4}\,(h\,{\rm Mpc}^{-1})^3$.
For lower number densities, it is advantageous to use a larger smoothing
scale to define the shifted field, as expected.
For comparisong, without reconstruction the full non-linear smearing at
$z=0$ leads to $\Sigma\simeq 10\,h^{-1}$Mpc, scaling as the growth factor
to higher redshift.
The horizontal dot-dashed line indicates the Silk scale, or the intrinsic
width of the acoustic peak, for our cosmology --- the observed width of the
acoustic peak is the quadrature sum of $\Sigma_{\rm Silk}$.

A different view is given in the lower panel of Figure \ref{fig:SigvsNR},
which shows how $\Sigma_{\rm eff}(z=0)$ depends on $R$ for different values
of $\bar{n}$.
Note the existence of an ``optimal'' smoothing scale, but that the minimum
is extremely broad.

These results show that, within the context of Lagrangian perturbation theory,
it is straightforward to understand the effects of noise in the density field
on the efficacy of reconstruction.  Reconstruction performs better when the
density field is determined from denser tracers, but the gains saturate at
$\bar{n}\sim 10^{-4}\,(h\,{\rm Mpc}^{-1})^3$.  For low density tracers it is
best to use a large smoothing scale to define the shifts, but the optimum is
very broad.

\acknowledgments

I would like to thank Joanne Cohn, Daniel Eisenstein, Yookyung Noh and
Nikhil Padmanabhan
for conversations and collaborations which significantly informed this work.
MW is supported by NASA and the Department of Energy.

\onecolumngrid

\bibliography{shotnoise}
\bibliographystyle{apsrev}

\end{document}